\documentclass[11pt,a4paper]{article}
\usepackage{amsmath,amssymb}
\usepackage{epsfig,graphicx}

\newcommand{\be}{\begin{equation}}
\newcommand{\ee}{\end{equation}}
\newcommand{\ba}{\begin{eqnarray}}
\newcommand{\ea}{\end{eqnarray}}
\newcommand{\n}[1]{\label{#1}}
\newcommand{\eq}[1]{Eq.(\ref{#1})}

\newcommand{\hh}{\, ,\hspace{0.5cm}}

\newcommand{\ve}{\varepsilon}

\begin{document}

\title{\bf Do Black Holes Exist?}
\author{Valeri~P.~Frolov\footnote{{\bf e-mail}: vfrolov@ualberta.ca},
\\
\small{\em Theoretical Physics Institute, University of Alberta,} \\
\small{\em Edmonton, AB, Canada,  T6G 2G7}
}
\date{}
\maketitle

\begin{abstract}
We discuss and compare definitions of a black hole based on the existence of event and apparent horizons. In this connection we present a non-singular model of a black hole with a closed apparent horizon and discuss its properties. We propose a massive thin shell model for consistent description of  particles creation in black holes. Using this model we demonstrate that for black holes with mass much larger than the Planckian one the backreaction of the domain, where the particles are created, on the black hole parameters  is negligibly small.
\end{abstract}

\section{Instead of introduction}

In 2015 we shall celebrate a hundredth anniversary of the Einstein General Relativity. One year later, in 2016, there will be the centenary of the publication of the famous paper by Schwarzschild solution, in which a static spherically symmetric solution of the Einstein equations was found. As it has been demonstrated later  this  solution describes a non-rotating static black hole. Much later, in 1963 Kerr discovered a generalization of the Schwarzschild, which describes a stationary rotating black hole. One may say that the black holes is a rather old subject of theoretical physics. In 1963 quasars were identified with very massive black holes. Now we believe that there are plenty of black holes in the Universe, both of stellar mass and supermasive black holes in the center of galaxies. The number of publications devoted to black holes is huge. Simple Google search  gives the  number of the documents which contain this notion  around 118,000,000.

One expects that after this hard work of many theoreticians and observational astrophysicists we know almost everything about black holes and their properties. However, this year papers were published with a claim that black holes do not exist. The first one is the paper by Stephen Hawking published on January 22, 2014
\cite{Hawk:14}. In which he wrote: "The absence of event horizons mean that
there are no black holes –- in the sense of regions from which light can't escape to infinity". This brief (4 pages) paper by Hawking received large publicity in the non-scientific journals and news-papers. In my talk at the conference I discussed properties of non-singular black holes with closed apparent horizons  \cite{FrVi:79,FrVi:81} -- the subject which closely related to Hawking publication. Such a model might be useful in the discussions of the information loss puzzle. At the same time, all the observable properties of such black holes remain practically the same as for the "standard" black holes, provided their mass is much larger than the Planckian one.

Later (already after the conference) two papers by Mersini-Houghton (the second one with H. Pfeiffer) appeared \cite{Mers_1,Mers_2}. Their main conclusion was that quantum effects prevent formation of stellar mass black holes in a massive star collapse. I strongly disagree with this conclusion and believe that these papers are wrong. For this reason I decided to add in this extended version of my talk, prepared for the publication in the Proceedings of the Quark-2014 meeting, some  remarks concerning this subject.

\section{"Physical black holes" vs "mathematical black holes"}

According to the standard definition, a black hole is a spacetime region from where  no
causal signals can escape to a distant observer. The event horizon is the black hole boundary. This definition can be put in mathematically exact form. This was done by Penrose \cite{Pen}.
According to this definition in order to decide whether a concrete astrophysical object is a black hole or not, one should know all its future evolution. In other words, one needs to live forever and during this infinite time to continue observations of a chosen suspicious object. This property is known as {\em teleological nature} of the event horizon. The above definition does not look realistic. However this definition is very convenient and allows one to prove a number of remarkable results concerning black hole properties. Let us call so defined  objects {\em mathematical black holes} (or briefly MBH).

Classical theorems on black hole  usually contain two types of assumptions. First, these are assumptions of regularity of the spacetime: no singularities in the black hole exterior, that can prevent predictability of the theory, no closed timelike curves and so on. Second type of assumptions concern the properties of the stress-energy tensor of the matter which is the source of the gravitational field. Usually it is assumed that one of the following,
so called, energy conditions is satisfied:
\begin{itemize}
\item Null energy condition, $\rho\ge 0$, $\rho+p\ge 0$;
\item Strong energy condition, $\rho+p\ge 0$;
\item Weak energy condition, $\rho+p\ge 0$, $\rho+3p\ge 0$.
\end{itemize}
Here $\rho$ and $p$ is the matter density and pressure. These conditions guarantee that the gravitational force is attractive and matter focuses light like a lens (with an astigmatism).
Well known results concerning the black hole properties are (see, e.g., \cite{Pen,MTW,HE,FrNo:98})
\begin{itemize}
\item Event horizon is generically null surface. Its generates being traced to  future do not have end points and the never leave it;
\item Black hole horizon has the topology of a sphere;
\item Black hole surface never decreases;
\item Stationary black holes are either axisymmetric (Kerr)
        or static (Schwarzschild);
\item Spacetime inside a black hole has a singularity.
\end{itemize}

However the definition of a MBH is not very useful for astrophysical applications when an observer, studying some object needs to decide whether he deals with a black hole or not. In this sense, the definition of MBH is unphysical. It is possible to give another definition of the black hole, which better reflects its physically important property, that the gravitational field in the vicinity of such an object is extremely strong. This definition is constructed as follows. One first defines a so called trapped surface. This is a compact smooth surface $B$ that has the property that both in- and out-going null surfaces, orthogonal to $B$, are non-expanding. A region inside $B$ is called a trapped region. A boundary of all trapped regions is called an {\em apparent horizon}. The existence of the apparent horizon means that gravitational field is so strong that it "traps" the light and particles in some space domain. To distinguish such an object, which has the apparent horizon, from MBH we call it a "physical black hole", or briefly  PBH.

The definitions of PBH and MBH are certainly different. Main difference is that PBH definition is more "practical". It does not require that an observer lives forever. However, generically the apparent horizon is not null and it does not serve as a "one way membrane", so that in principle some information can be obtained from PBH "interior".

In classical physics, in the absence of exotic matter, when the energy conditions are satisfied, there is a correlation between existence of the event and apparent horizons. Namely, if the apparent horizon exists it always lies inside (or coincides with) the event horizon \cite{HE}. In this sense existence of a PBH implies the existence of a MBH.

\section{Quantum effects}

In quantum physics energy conditions can be violated. I believe that Markov was the first who mentioned this possibility in the connection with black holes \cite{Mark:74}. The gravitational field of a black hole becomes stationary soon after the collapse. For simplicity we assume that a  black hole is static spherically-symmetric and discuss some features of the particle creation in its gravitational field.

Let us consider first some static field (not necessary gravitational field) and denote its strength by $\Gamma$. We denote by $g$  a coupling constant of interaction of particles of mass $m$  with it.
There are virtual pairs of such particles in a vacuum and the probability for the particles of the pair to be a distance $\ell$ from one the other is
\be
P\sim e^{-S}\hh S\sim \ell/\ell_{m}\, .
\ee
Here $\ell_m=\hbar/mc$ is the Compton wavelength of mass $m$. A virtual pair will not disappear only if the work of the external force on separated particles would be sufficient and provide the particles with the energy that make them real (that is, "moves" their momenta to the mass shell). Such long living ("real") particles form the quantum radiation. The corresponding condition is
\be
g\Gamma \ell=2mc^2\, .
\ee
Thus one can estimate the probability of the pair creation by the external field as follows
\be\n{gen}
P\sim e^{-S}\hh S\sim {2m^2 c^3\over \hbar g\Gamma}\, .
\ee

The best known example of such an effect is charged particles (charge $e$) creation in the static homogeneous electric field of the strength $E$. The famous Schwinger \cite{Schwinger} result  is
\be\n{schw}
P=VT {e^2E^2\over 8\pi^2 \hbar^2 c}\exp\left(-{\pi m^2 c^3\over \hbar |eE|}\right)\, .
\ee

It is easy to see that the estimation (\ref{gen}) correctly (up to numerical factor of the order of one) reproduces the exponent in the expression (\ref{schw}).
The presence of $VT$ in the prefactor indicates that the production of particles takes place uniformly over the total volume $V$ and permanently in time  $T$. The energy required for the particle production is provided by the electric field. Suppose  we switch on an electric capacitor of the volume $V$ and the electric field $E$ for time $T_0$ and after switch it off. For large $T_0$ the total number of created particles can be written as follows
\be\n{num}
N=C+ n T_0\, .
\ee
The quantity $C$ depends of details of the switching -on and -off procedures and $n$ is the number of particles per a unit time produced by the field.

Is it possible to "explain" the Schwinger effect by saying that all the particles are produced only  when the field was time-dependent, that is during the switching-on the field, so that later the created particles simply leak out? Evidently, this is impossible because such an "explanation" violates the causality. Really, one can change the duration $T_0$ of the constant field phase (and hence the total number of created particles) without modifying the switching-on process.

Detailed study of the Schwinger process allows one to arrive to the following conclusions:
\begin{itemize}
\item Particles of the pair are created at spacelike distance from one the other;
\item After creation they move in the opposite directions with the same value of constant four-acceleration $a=|eE|/m$;
\item During the process of creation there exists vacuum polarization current $\langle j^{\mu}\rangle$ which is spacelike.
\end{itemize}

Particle creation by black holes is in many aspects similar to the Schwinger process. Let us use relation (\ref{gen}) to estimate the rate of particle production by a static gravitational field. Namely, we put $g=m$ and identify $\Gamma$ with the surface gravity of the black hole
\be
\kappa={GM\over r_g^2}={c^4\over 4GM}\, .
\ee
Relation (\ref{gen}) takes the form
\be\n{grav}
P\sim e^{-E/\Theta}\, .
\ee
Here $E=mc^2$ is the energy of the particle and
\be
\Theta={\hbar c^3\over 8GM}\, .
\ee
This estimation correctly reproduces the Hawking result if instead of $\Theta$ one uses the Hawking temperature
\be
\Theta_H={\hbar c^3\over 8\pi G M}\, .
\ee
which differs from $\Theta$ by a factor $\pi$.
As a result of the Hawking radiation the black hole loses its mass
\be
\dot{M}\sim -C \left( {m_{Pl}\over M}\right)^2\, .
\ee
Here $C$ is dimensionless coefficients that depends on the number and properties of the emitted particles. The lifetime of an isolated black hole is
\be
T_{evap}\sim t_{Pl} \left( {M\over m_{Pl}}\right)^3\, .
\ee
If the black hole is isolated and there is no matter falling into it, its mass decreases and the temperature $\Theta_H$ (and hence the rate of energy emission) grows.

However, one can support the constant value of $\Theta_H$ for arbitrary long time $T_0$ by feeding  the evaporating black hole with some accreting matter, that compensates the mass loss. If $T_0$ is much larger than both, the time  of the black hole formation, $T_{collapce}$, and $T_{evap}$  total number of created particles can be written in the form (\ref{num}). The conclusion is that besides particles created during the collapse, the number of which it completely determined by the collapse details and usually is very small, there exists permanent production of  particles by the constant (or slowly evolving) gravitational field of the black hole. Arguments similar to those discussed for the Schwinger effect show that one cannot (without violation of causality) explain the production of the particles by a black hole by saying that all these particles have been created during the formation of the black hole (that is at the stage of the collapse) and  their "leakage" out the domain where they have been formed is the origin of the Hawking flux. This simple observation demonstrates that a similar point of view, adopted in \cite{Mers_1,Mers_2}, is wrong.

The particle production by black holes is in many aspects similar to the Schwinger process and it possesses the properties:
\begin{itemize}
\item Particles are created in pairs;
\item Particles are created at spacelike distance. One particle is created outside the black hole, while the other is created inside the horizon, where its energy is negative;
\item Those of the positive energy particles, that penetrate the potential barrier and reach infinity, form the Hawking radiation;
\item During the process of the pair production there exist vacuum polarization flux of negative energy through the black hole horizon. Namely this flux reduces the mass of the black hole.
\end{itemize}

\section{Massive thin shell model}

Let us describe a simple model illustrating the above formulated features of the quantum emission of a black hole. Massless particles production dominates in this process. The outgoing energy flux can be effectively described by a properly chosen null fluid. The conservation law requires that this radiation is accompanied by the negative energy flux through the horizon, which we also approximate by the null fluid. In order to make a model consistent one needs to assume that between the two regions with pure outgoing and pure incoming fluxes there exists a transition region, corresponding to the domain where the particle are created. We assume that this region is narrow and approximate it by a massive thin shell \cite{Hayw:06,Fr:14}. Let us demonstrate consistency of such a model and show that for slowly evolving black holes the backreaction of the shell on the metric is negligibly small.

To describe the metric near a horizon we use the Vaidya solution which we write in the form
\be
dS_{\ve}^2=-F_{\ve}dZ_{\ve}^2 -2\ve dZ_{\ve} dr+r^2 d\omega^2\, ,\  F_{\ve}=1-{2M_{\ve}(Z_{\ve})\over r}\, .
\ee
Here $d\omega^2$ is the metric on a unit sphere and $\ve=\pm 1$. For $\ve=+1$ the coordinate $Z_{+}$ is a retarded null time $U$, while for $\ve=-1$, $Z_{-}$ is an advanced null time $V$. The Ricci tensor calculated for this metric is
\be
R^{\ve}_{\mu\nu}=-{2\ve\over r^2}{dM_{\ve}\over dZ_{\ve}} Z_{\ve;\mu}Z_{\ve;\nu} \, .
\ee
If ${dM_{\ve}/dZ_{\ve}}<0$ then according to the Einstein equations  the corresponding stress-energy tensor for $\ve=+1$
\be
T^{+}_{\mu\nu}=-{1\over 4\pi r^2}{dM_+\over dU} U_{,\mu}U_{,\nu}
\ee
describes outgoing null fluid with positive energy density. The corresponding
stress-energy tensor for $\ve=-1$
\be
T^{-}_{\mu\nu}={1\over 4\pi r^2}{dM_-\over dV} V_{,\mu}V_{,\nu}
\ee
describes incoming null fluid with negative energy density.

Consider a spherical surface $\Gamma$ outside the horizon described by the equation
\be
Q(r,Z)=r-R(Z)=0\, ,
\ee
and assume that outside this surface the metric is $dS_+^2$ while inside it is $dS_-^2$. We also assume that the null coordinate $V$ in the inner region is chosen so that on $\Gamma$ it coincides with the corresponding value of $U$
\be
V|_{\Gamma}=U|_{\Gamma}=Z\, .
\ee
The intrinsic 3-geometries induced by the metrics $dS_{\pm}^2$ on $\Gamma$ are
\be
d\sigma_{\ve}^2=- \left[1-{2M_{\ve}(Z)\over R}+2\ve R'\right] dZ^2+R^2 d\omega^2\, .
\ee
Here $R'=dR/dZ$. Both metrics are identical when the following condition is satisfied
\be
M_+(Z)-M_-(Z)=2RR'\, .
\ee
We assume that this condition is satisfied and write $M_{\pm}$ in the form
\be
M_{\ve}(Z)=M(Z)+\ve RR'\, ,
\ee
so that the induced metric on $\Gamma$ (which is the same for $\ve=\pm1$) is
\be
d\sigma^2=- (1-{2M(Z)\over R}) dZ^2+R^2 d\omega^2\, .
\ee

We choose a special form of $\Gamma$ and put $R=2M(Z)(1+w)$, where $w$ is a dimensionless  positive small parameter. Then calculations give the following expressions for the non-vanishing components of $K^{\ve}_{\hat{i}\hat{j}}$
\be
K^{\ve}_{\hat{t}\hat{t}}=A+\ve B\, ,\
K^{\ve}_{\hat{\theta}\hat{\theta}}=-2wA+\ve B\, ,\
K^{\ve}_{\hat{\phi}\hat{\phi}}=-2wA+\ve B\, ,
\ee
where
\be
A={1\over 4M\sqrt{w(1+w)^3}}\hh B={M'\sqrt{1+w}\over \sqrt{w} M}\, .
\ee

The jumps of the extrinsic curvature at $\Gamma$ determine parameters (mass and pressure) of the shell. These jumps are
\be
[K_{\hat{t}\hat{t}}]=[K_{\hat{\theta}\hat{\theta}}]=[K_{\hat{\phi}\hat{\phi}}]=2B\, .
\ee
One can interpret the corresponding distribution of matter as being connected with a region of particle creation. In any case for a slow change of the black hole parameters, that is when $|M'|\ll 1$, the influence of this matter on the black hole geometry is proportional $|M'|$ and hence is extremely small and can be neglected.

Let us focus on the flux that reached the infinity at some retarded time $U$ and trace it back in time. In this inverse motion it comes close to the horizon and keeps going along it. In the above described model when this beam of flux reaches the domain of particle creation, represented by the massive shell, it meets the negative energy flux and `annihilates' it. What would happen if such a flux is not a result of the vacuum decay, but it is the result of "leakage" of particles that have been created during the collapse. These particles must have very high energy in order to escape the collapsing matter just prior to the moment, when it crosses the gravitational radius, and produce later, at time $U$ energy flux $\varepsilon_{Hawking}$.
The corresponding energy density $\varepsilon_{init}$ of the null fluid flux at the stage of the collapse must be exponentially large
\be
\varepsilon_{init}\sim \exp(\kappa U) \varepsilon_{Hawking}\,
\ee
In such a scenario  one would arrive to a wrong conclusion that back-reaction of these `prior' fluxes may prevent the black hole formation. Namely this mistake was made in the  papers \cite{Mers_1,Mers_2}.

\section{Spacetime with a closed apparent horizon}

\subsection{Limiting curvature conjecture}

Quantum effects in black holes violate null energy conditions, so that logically the existence of PBH does not imply the existence of MBH. One of the options is that the spacetime inside the black hole is regular and the apparent horizon is closed. Such a model was first proposed in \cite{FrVi:79,FrVi:81}.
In particular, the abstract of the  paper \cite{FrVi:81} contains the following statement: "For large M it is conjectured that the event horizon does not form, and the apparent horizon is closed. An object forms, possessing the observable properties of a black hole, but living a finite time."
Later similar models were discussed in many publications (see e.g. \cite{Hayw:06,Fr:14,RoBe:83,Solo:99,Grum:03,Grum:04,Anso:08,BaMaMo:13,RoVi:14,KaYo,Bard}). Let us illustrate some properties of regular black holes with a closed apparent horizon by a simple model, following the paper \cite{Fr:14}.

We assume that the initial mass $M$ of the black hole is much larger than the Planckian mass $m_{Pl}$, so that the black hole is a classical object. We also assume that its gravitational field during all the evolution (including the final state of the evaporation) as well in the black hole interior is described by an effective metric tensor $g_{\mu\nu}$. In the domain where quantum corrections are small it obeys the Einstein equations. At the initial stage of the evaporation one can use quasi-classical description, so that the Hawking process results in the positive energy flux of created particles to infinity and (in accordance with the conservation law) by the negative energy flux through the horizon. The latter decreases the mass of the black hole. The main contribution to this process is due to massless fields, so that we describe the energy flux to future null infinity by properly chosen null fluid stress-energy tensor. We assume that the negative energy flux through black hole horizon into its interior can also be approximated by the null fluid with negative energy.

In the spacetime domain, where the curvature becomes comparable with the Planckian one, the classical Einstein equations should be modified as a result of the effects of the vacuum polarization and intensive quantum particle creation. For the black hole with the gravitational radius $r_S=2M$ the curvature reaches the Planckian value at $r=r_0$ which can be found from the equation
\be
{r_S\over r_0^3}\sim l_{Pl}^{-2}\, .
\ee
In the spacetime domain where
\be\n{r0}
r<r_0=r_S (l_{Pl}/r_S)^{2/3}
\ee
the curvature calculated for the classical solution exceeds the Planckian one.

\subsection{Modified Vaidya metric}

Suppose a black hole is formed by spherical collapse of null fluid. The formation of such a black hole is described by the Vaidya solution
\be\n{vaidya}
dS^2=-F dV^2+2dV dr+r^2 d\omega^2\, ,\ F=1-{2M(V)\over r}\, .
\ee
It is easy to check that
\be
(\nabla r)^2=F \, ,
\ee
and hence the apparent horizon is located at the surface
\be
r=2M(V)\, .
\ee

We assume that collapse starts at $V=V_0<0$ and before this time the mass $M$ vanishes. The apparent horizon is formed at $r=0$ at time $V_0$ . It is easy to show that if only the apparent horizon crosses $r=0$, then the spacetime cannot be regular there and the curvature singularity is formed. Indeed, the square of the Riemann tensor
\be
{\cal R}^2\equiv R_{\mu\nu\alpha\beta}R^{\mu\nu\alpha\beta}={48 M_-(V)^2\over r^6}\, .
\ee
is divergent at $r=0$.

We assume that collapse ends at time $V=0$, when the black hole mass reaches some value $M_0$, and after this only Hawking radiation changes the black hole parameters.
The negative energy flux through the horizon, that accompanies the  particle creation, reduces the mass of the black hole. We use the same Vaidya solution (\ref{vaidya}) to describe it. But now $M(V)$ is a decreasing function of $V$. The rate of change of the mass can be approximated by the formula
\be\n{hawk}
{dM\over dV}=-C {m_{Pl}\over t_{Pl}}\left({m_{Pl}\over M}\right)^2\, .
\ee
In fact $C$ also depends on time, since the number of different species of particles that can be emitted grows with increase of the black hole temperature. In what follows to simplify the formulas we neglect this dependence. Moreover, we simply put  $C=1/3$.

We assume that $M_0\ge m_{Pl}$ so that during the initial stage of the evaporation the region \eq{r0}, where the classical Einstein equations should be modified,
is located deep inside the black hole. The modification is also  required in order to describe the final stage of the evaporation. In these domains the metric is expected to be quite different from its `classical form'. Since we do not have a reliable theory of gravity valid at the Planckian scales we use special ansatz for the metric in these domains. We specify a model so that it satisfies the limiting curvature principle \cite{Markov:82,Markov:84}.
This prescription certainly contains an ambiguity. However even consideration of such simple models is quite instructive and allows one to study their robust predictions.

To describe an evaporating black hole with a regular interior we use a modified Vaidya metric  \cite{Hayw:06,Fr:14}.
This metric has the same form as \eq{vaidya} with a modified function $F$
\be
F=1-{2M(V)r^2\over r^3+2M(V) b^2}\, .
\ee
Here $b$ is the cut-off parameter of the order of the Planck length. For simplicity we put $b=l_{Pl}$. For $M\gg b$ and $r\gg r_0$, where $r_0$ is given by \eq{r0} the modified metric remains practically the same as earlier. However for $r\ll r_0$ it takes the form
\be
F\sim 1-(r/b)^2\, .
\ee
This implies that the geometry near $r=0$ is regular and  $r=0$ is a regular timelike line. Denote
\be
q={2M b^2\over r^3}\, ,
\ee
then the curvature invariants for the modified Vaidya metric are of the form
\be
{\cal R}^2={12 q^2(1-4q+18q^2-2q^3+2q^4)\over b^4 (1+q)^6}\, .
\ee
For small $q$ one has
\be
{\cal R}^2\sim {12 q^2\over b^4}={48 M^2\over r^6}\, .
\ee
In the limit $b\to 0$ one restores the expressions for the original Vaidya metric. In the opposite case (when $q\to \infty$) one has
\be
{\cal R}^2\sim 24 b^{-4}\, .
\ee
These relations show that the curvature of the modified Vaidya metric is limited and its maximal value is of the order of $b^{-2}$.

\subsection{Apparent horizon}

In what follows it is convenient to use dimensionless quantities by choosing $b$ (the cut-off parameter which enters in the modified Vaidya metric) as a natural scale parameter. We denote
\be
v=V/b\hh\mu=M/b\hh\rho=r/b\, .
\ee
We also denote by prime the dimensionless time derivative: $(\ldots)'=d(\ldots)/dv$.

\begin{figure}[tbp]
\centering
\includegraphics[width=6cm]{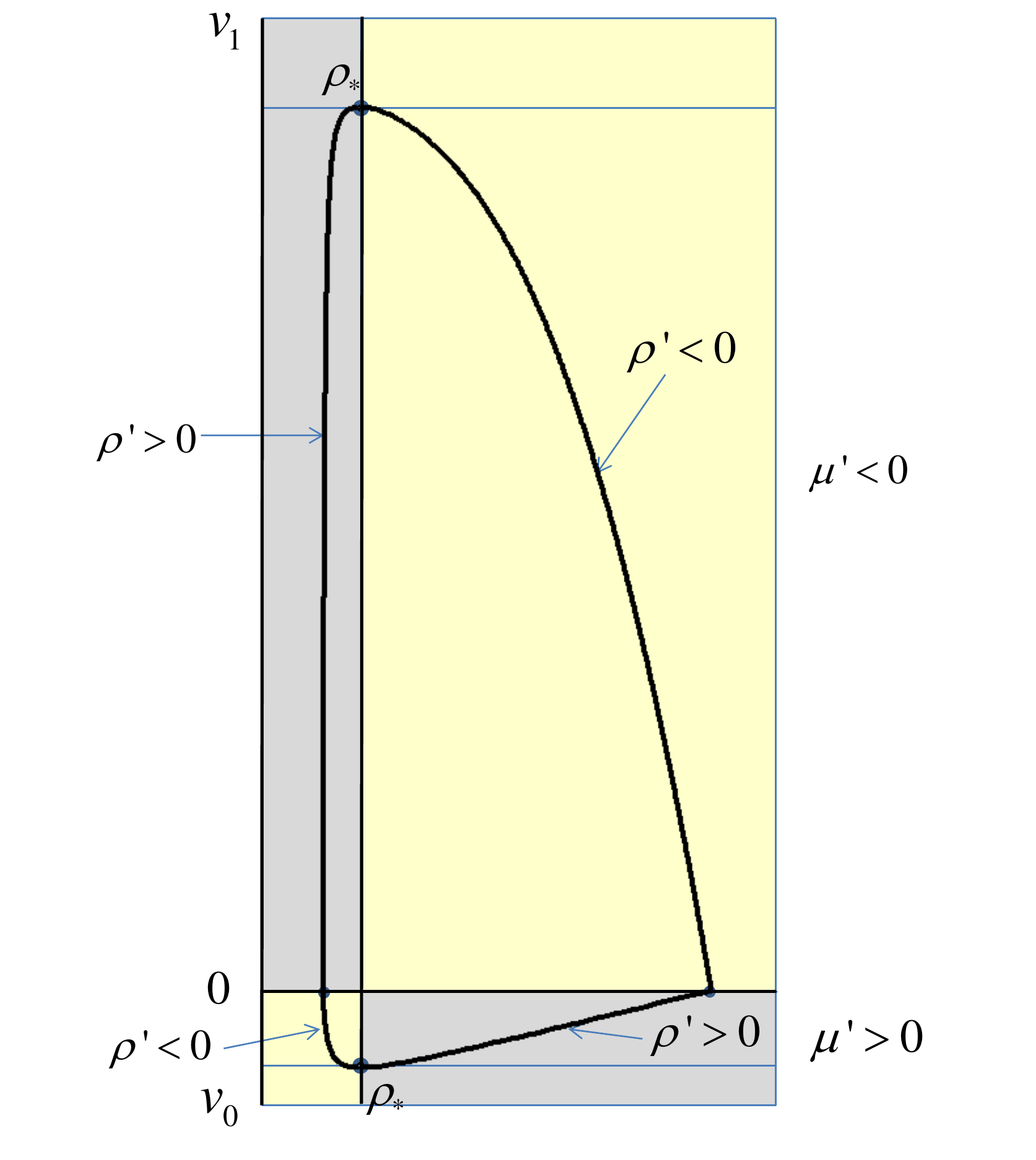}\hfill
  \caption{Apparent horizon structure.
\label{Fig_1}}
\end{figure}

We specify the function $\mu(v)$ as follows: it vanishes before $v_0$ (the moment when the collapse begis), linearly grows during the time interval $(v_0,0)$ and reaches its maximum value $\mu_0$ at $v=0$.  After this it monotonically decreases according to the law
\be\n{mu}
\mu^3(v)=\mu_0^3-v\, ,
\ee
until it vanishes at $v_1=\mu_0^3$ (the end of the evaporation).
After this the mass is identically zero.

For this scenario the apparent horizon is represented by a closed line in $(v,\rho)$ plane. It appears and disappears at the moments $v^*_{-}$ and $v^*_{+}$, respectively ( $v_0<v^*_{-}<v^*_{+}<v_1$). At the moments $v^*_{\pm}$ the inner and outer brunches of the apparent horizon coincide
\be
\rho_{\pm}(v^*_{\pm})=\rho_*\, .
\ee
Figure~\ref{Fig_1} shows the structure of the apparent horizon for this model.

This is a special case of a general class of non-singular black holes models with a closed apparent horizon which was first discussed in the paper \cite{FrVi:79,FrVi:81}. For large $M/m_{Pl}$ this is an example of a PBH which does not have the event horizon and hence it is not a MBH. The main feature of such an object is the following: during all the time of its evaporation the gravity at its outer surface is so strong that it forces even out-going radial null rays to propagate to the center. As a result, an external observer during all this time cannot see anything in the PBH abyss. However, after the evaporation ends and the apparent horizon disappears, the information from the PBH's interior might return to an external observer.

\begin{figure}[tbp]
\centering
\includegraphics[width=6cm]{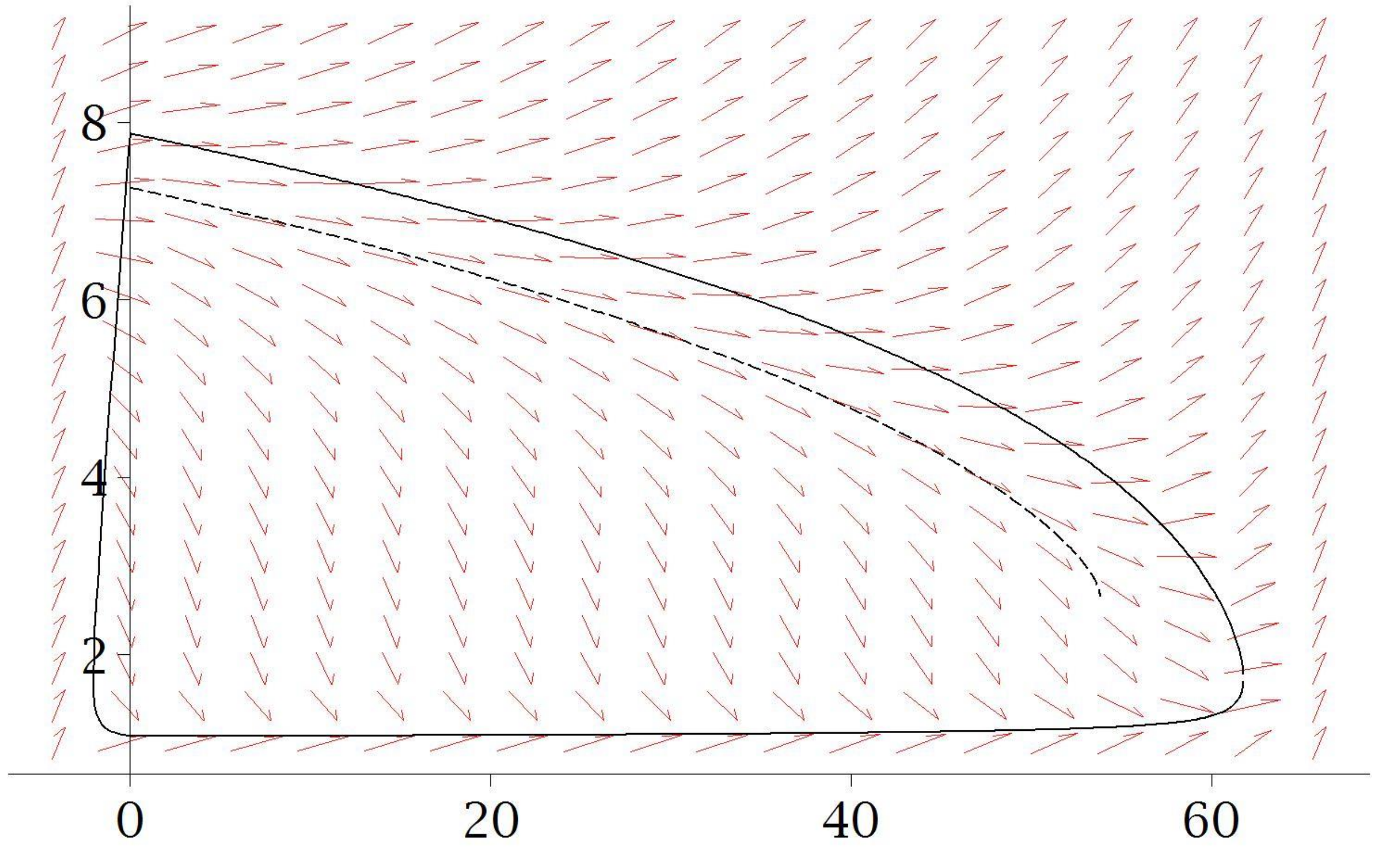}
  \caption{`Outgoing' radial null rays in the modified Vaidya metric. The horizontal axis is $v$ and the vertical one is $\rho$. The vector field $l^{\mu}=(1,f/2,0,0)$ tangent to the outgoing null rays is shown by small arrows. The plot is constructed for $\mu_0=4$ and $v_0=-3$. A closed solid line shows a position of the apparent horizon. A dashed line is a quasi-horizon.\label{Fig_2}}
\end{figure}

The Figure~\ref{Fig_2} illustrates the propagation of the radial out-going null rays in the above model. It clearly shows that, in fact, light can escape from a narrow region inside the apparent horizon. At the same time there exist another surface which plays the role of "no-escape membrane". It is shown by a dashed line at the Figure~\ref{Fig_2}. This surface is a separatrix. It separates two families of radial out-going null rays, those that escape to infinity during the evaporation process and those that are trapped during the time of the evaporation and can escape to infinity only after the end of the evaporation. This surface is defined by the equation
\be
2\partial_V F+F\ \partial_{r}F=0\, .
\ee
It is possible to show \cite{Fr:14} that for slow evaporation, $|\mu'|\ll 1$, it is practically null.  We call this separatrix a {\em quasi-horizon}.

The described picture is valid only if the  collapsing object has the mass much larger than the Planckian mass.
It is possible to show that the apparent horizon is not formed  if the mass a collapsing object is smaller than $\sqrt{3}$. One may expect that the existence of the mass gap for PBH formation in the gravitational collapse is a common feature of models of regular black holes with a closed apparent horizon \cite{FrVi:79,FrVi:81}.

\section{Discussion}

It should be emphasized that the described model is certainly over-simplified. The effective metric $g_{\mu\nu}$ is used to describe the gravitational field  in the domain where the curvature reaches the Planckian value. However, we do not know how the gravitational equations are modified in this domain. Instead of this one uses the principle of the limiting curvature \cite{Markov:82,Markov:84} and specifies the model to satisfy this principle. In the described  model with a closed apparent horizon both the matter, originally forming the black hole, and the quantum radiation, formed inside the black hole and carrying negative energy, are compressed in a tiny region near the inner horizon. They have  very high (super Planckian) density. This means that the non-linear quantum interaction between particles as well as their interaction with gravity becomes important. One can also expect, that quantum fluctuations of the metric are important and that they can "smear" the position of the inner horizon.

In the proposed model there is a strong blue-shift effect for rays propagating in the vicinity of the inner horizon  \cite{Fr:14,BoFr:86}. These rays are focused and produce exponentially thin beams of ultra-high frequency. This phenomenon has common features with  a well known transplanckian problem in black holes \cite{Jaco:91} and in the inflating universe \cite{MaBr:01}. One can expect mutual annihilation of particles propagating along the inner horizon. This effect in some sense is inverse to the Hawking effect of particles creation. Let us also mention that if the low energy gravity is an emergent phenomenon, then for proper description of the matter in the inner `core' one needs to use the fundamental background theory.

Main feature of the presented model is that after the complete evaporation of the black hole  the information accumulated inside the black hole  becomes available to an external observer. This opens a possibility of the restoration of the unitarity in evaporating `black holes'. However in order to reinstate the information lost during the collapse of mass $M$ the remnant at the last stage of evaporation must emit $\sim (M/m_{Pl})^2$ quanta, so that each quantum has typical energy $m_{Pl}^3/M^2$. Estimations show that this process may take quite long time $\sim t_{Pl} (M/m_{Pl})^4$ \cite{Ah,Car,Pres}.
All these problems are connected with unknown properties of the quantum gravity, the theory which does not exist at the moment. However, a proposed model of a `black hole' with a closed apparent horizon allows one to address interesting questions and test  possible new ideas.

Let us emphasize that the effect of quantum evaporation (and all the related problems) is important only for black holes of small mass. In the "standard" theory of gravity the lifetime of black holes of mass greater than $10^{15}$g is larger than the age of the Universe. Suppose that such a black hole has both the apparent and event horizon. It is easy to estimate the radial distance between them which arises  as a result of quantum evaporation
\be
|r_{EH}-r_{AH}|\sim {l_{Pl}^2\over r_{AH}}\, .
\ee
This means that if the mass of the black hole is much larger that the Planckian one, this difference is much smaller than the Planckian length. One can expect that due to the fluctuations of the metric it is simply impossible to measure this difference in principle.

If the accretion rate of surrounding matter on an astrophysical black hole is  high the size of the apparent horizon is different from the size of the event horizon.  But, as we have already emphasized, the position of the event horizon cannot be measured \cite{Vis}. In such a situation the best that one can observe (say, by the Event Horizon Telescope or by observations of black-hole--black-hole collisions by the gravitational waves detectors like LIGO) is  the size and shape of the apparent horizon. In any case for astrophysical black holes quantum effects are negligibly small and these tiny effects cannot change the standard  picture of the gravitational collapse. The opposite opinion, presented in \cite{Mers_1,Mers_2}, is based on mistakes and is wrong.

The problem of existence of the event horizon and the related problem of the information loss are very important in application to small black holes. However they will become of "practical importance" only when such small black holes will be found, say either as primordial black holes or as micro black holes, produced in colliders or by cosmic rays.

\medskip

{\bf Acknowledgments}

\medskip

The author is grateful to the Natural Sciences and Engineering
Research Council of Canada and the Killam Trust for their financial support.

\end{document}